\documentclass[a4paper,10pt,twoside]{cpc-hepnp}
\usepackage{multicol}
\usepackage{graphicx}
\usepackage{booktabs}
\usepackage{amssymb,bm,mathrsfs,bbm,amscd}
\usepackage[tbtags]{amsmath}
\usepackage{lastpage}

\usepackage{float}
\usepackage{multirow}
\usepackage{slashed}
\usepackage{tabularx}
\usepackage{hyperref}
\usepackage{subfig}
\usepackage{lineno}
\usepackage{color}
\usepackage{soul}
\begin{document}
\title{Spin Flavor Spectral Splits of Supernova Neutrino Flavor Conversions}

\maketitle

\begin{center}
{\bf Ziyi Yuan~$^{a,~b}$}~\footnote{Email: ziyiyuan@whu.edu.cn},
{\bf Yu-Feng Li~$^{b,~c}$}~\footnote{Email: liyufeng@ihep.ac.cn (corresponding author)},
{\bf Xiang Zhou~$^{a}$}~\footnote{Email: xiangzhou@whu.edu.cn (corresponding author)},
\\
\vspace{0.2cm}
{$^a$School of Physics and Technology, Wuhan University, Wuhan 430072, China}\\
{$^b$Institute of High Energy Physics, Chinese Academy of Sciences, Beijing 100049, China}\\
{$^c$School of Physical Sciences, University of Chinese Academy of Sciences, Beijing 100049, China}
\end{center}

\begin{abstract}
The supernova neutrino flavor evolution in the presence of the non-trivial neutrino magnetic moment and strong magnetic field is numerically derived using the two-flavor and single-angle approximation. The novel properties of collective neutrino oscillations are studied and distinct patterns of flavor and spin-flavor spectral splits are presented. Finally we also discuss how the neutrino magnetic moment affects the observable supernova neutrino energy spectra.
\end{abstract}


\begin{multicols}{2}
\section{Introduction}

The first observation of neutrinos from the Supernova (SN) 1987A explosion~\cite{Hirata:1987hu,Bionta:1987qt,Alekseev:1988gp} represents an important milestone for both neutrino physics and neutrino astronomy, which has triggered intensive interest for high-statistics detection of SN neutrinos for the next Galactic SN explosion using large neutrino detectors~\cite{Abe:2016waf,Abe:2021vhc,An:2015jdp,Abi:2020lpk}.
The observation of SN neutrinos can not only help us to establish the SN explosion mechanisms~\cite{Janka:2012wk}, but also provide an excellent laboratory to study the neutrino properties in the extreme environment~\cite{Mirizzi:2015eza}.

After the establishment~\cite{Zyla:2020zbs} of solar neutrino flavor conversions with the Mikheyev-Smirnov-Wolfenstein (MSW) matter effects~\cite{Mikheev:1986gs,Mikheev:1986wj,Wolfenstein:1977ue}, it has been realized that the matter effects may have significant impact on the flavor conversion of SN neutrinos~\cite{Mirizzi:2015eza}. The ordinary MSW flavor conversions may occur in the outer layers of the SN mantle or envelope, whose main features depend on the SN density profile and the neutrino mass ordering~\cite{Dighe:1999bi,Lunardini:2003eh}. Moreover, the neutrino coherent forward scattering on the dense neutrino background may be important in the SN environment~\cite{Fuller,Notzold:1987ik,Pantaleone:1992eq},
and non-linear collective neutrino oscillations, including the synchronized and bipolar ones, would arise in the presence of the neutrino self-interaction potential~~\cite{Duan:2010bg}, in which the neutrinos may have almost complete flavor conversions.
An interesting observable phenomenon for collective neutrino oscillations is the existence of the spectral splits in the SN neutrino energy spectra of different neutrino flavors~\cite{Duan:2006an,Fogli:2007bk,Duan:2008za,Dasgupta:2009mg}.

In this work we shall provide a formalism on the neutrino flavor evolution equation in the presence of the non-trivial neutrino magnetic moment and strong magnetic field, and study novel properties of the collective neutrino oscillations. By using the two-flavor and single-angle approximation, we are going to make a numerical calculation of the neutrino evolution equation inside the dense SN medium. We shall also study the patterns of neutrino flavor and spin-flavor spectral splits for both normal ordering (NO) and inverted ordering (IO) of the neutrino mass spectrum and discuss how the neutrino magnetic moment affects the observable neutrino energy spectra.

The remaining part of this work is organized as follows. In Sec.~2, we present our calculation framework for the neutrino flavor evolution equation in the presence of the neutrino magnetic moment and strong magnetic field. Then, results of the numerical calculations for the evolution equation are given in Sec.~3, and different patterns of the collective neutrino oscillations and energy spectral splits will be provided. Finally, we summarize the main results and conclude in Sec.~4.

\section{Calculation Framework}

First of all, in this section we shall introduce the calculation framework of this study. We start with the bulb model of supernova neutrino emission, in which the proto-neutron star emits neutrinos and antineutrinos uniformly and isotropically outward in all directions~\cite{Duan:2006an}.
The emitted neutrino is characterized by its momentum $\mathbf{q}$ ($|\mathbf{q}|=E$),
with the incident polar angle $\theta_{\mathbf{q}}$, azimuthal angle $\phi_{\mathbf{q}}$ and radial position $r$ in the SN medium.
The differential neutrino number density $\operatorname{d}n_{\nu_\alpha}(\mathbf{q})$ is then given by
\begin{equation}
    \operatorname{d}n_{\nu_\alpha}(\mathbf{q})=\frac{L_{\nu_\alpha}}{4 \pi^2 R_{\nu_\alpha}^{2}\langle E_{\nu_\alpha}\rangle} \phi_{\nu_\alpha}(E)\operatorname{d}(\cos{\theta_\mathbf{q}})\operatorname{d}\phi_\mathbf{q}\ ,
\end{equation}
where {$R_{\nu_\alpha}=50\;\operatorname{km}$} is the radius of the neutrino sphere, $ L_{\nu_\alpha}=10^{51}\;\operatorname{erg/s}$ is the $\nu_{\alpha}$ luminosity.
$\langle E_{\nu_\alpha}\rangle$ is the average energy of the $\nu_{\alpha}$ flux, which is taken as
$\langle E_{\nu_e}\rangle=10\;\operatorname{MeV},\;\langle E_{\overline{\nu}_e}\rangle=15\;\operatorname{MeV},\;\langle E_{\nu_x}\rangle=\langle E_{\overline{\nu}_x}\rangle=24\;\operatorname{MeV}$ for $\nu_e$, $\overline{\nu}_e$, $\nu_x$ and $\overline{\nu}_x$, respectively, where $\nu_x$ and $\overline{\nu}_x$ denote all the non-electron flavors $\nu_\mu$, $\nu_\tau$, and $\overline{\nu}_\mu$, $\overline{\nu}_\tau$ respectively.
According to the geometry of the neutrino bulb model and by assuming single angle approximation where the neutrino evolution history is trajectory independent~\cite{Duan:2006an,Fogli:2007bk}, the neutrino flux at the radial position $r$ can be obtained using an additional geometric factor $D(r)$
\begin{eqnarray}
        D(r)&=& \left. \int_{\cos{\theta_{\rm max}}}^{1}\operatorname{d}(\cos{\theta_\mathbf{q}})(1-\cos{\theta_\mathbf{q}}\cos{\theta_\mathbf{p}})\right|_{\cos{\theta_\mathbf{p}}=1} \nonumber\\
        &=&\frac{1}{2}\left[1-\sqrt{1-\left(\frac{R_\nu}{r}\right)^{2}}\right]^2.
\end{eqnarray}
The normalized neutrino spectra $ \phi_{\nu_\alpha}(E)$ can be calculated as
\begin{equation}\label{initial spectra}
    \phi_{\nu_\alpha}(E)=\frac{2\beta_{\nu_\alpha}}{3\zeta}\frac{(\beta_{\nu_\alpha} E)^2}{e^{\beta_{\nu_\alpha} E}+1},
\end{equation}
where $\zeta\approx 1.202$, and $\beta_{\nu_e}=0.315\;\rm{MeV}^{-1},\;\beta_{\overline{\nu}_e}=0.210\;\rm{MeV}^{-1},\;\beta_{\nu_x}=\beta_{\overline{\nu}_x}=0.131\;\rm{MeV}^{-1}$.

The flavor evolution of the neutrino states can be calculated using the density matrix formulation
\begin{equation}\label{evoltion equation}
    \operatorname{i}\dot\rho=\left[H,\;\;\rho\right],
\end{equation}
where the Hamiltonian $ H $ includes four different components:
\begin{equation}\label{hamiltonian}
H(r,\mathbf{q})=H_{\rm vac}(E)+H_{\rm mat}(r)+H_{\rm self}(r,\mathbf{q})+H_{\rm mag}(r)\,,
\end{equation}
and the density matrix $\rho$ in the flavor basis of both neutrinos and antineutrinos ($\nu_{e}$, $\nu_{x}$, $\overline{\nu}_{e}$, $\overline{\nu}_{x}$) can be written as:
\begin{equation}
    \rho=
    \begin{pmatrix}
        \rho_{e e} & \rho_{e x} & \rho_{e \bar{e}} & \rho_{e \bar{x}} \\
        \rho_{x e} & \rho_{x x} & \rho_{x \bar{e}} & \rho_{x \bar{x}} \\
        \rho_{\bar{e} e} & \rho_{\bar{e} x} & \rho_{\bar{e} \bar{e}} & \rho_{\bar{e} \bar{x}} \\
        \rho_{\bar{x} e} & \rho_{\bar{x} x} & \rho_{\bar{x} \bar{e}} & \rho_{\bar{x} \bar{x}} \\
    \end{pmatrix}\,.
\end{equation}

The first component $H_{\rm vac}(E)$ in Eq.~\ref{hamiltonian} is the effective vacuum Hamiltonian:
\begin{equation}\label{H_vac1}
    H_{\rm vac}=\omega
    \begin{pmatrix}
        -\cos2\theta & \sin2\theta & 0 & 0 \\
        \sin2\theta & \cos2\theta & 0 & 0 \\
        0 & 0 & -\cos2\theta & \sin2\theta \\
        0 & 0 & \sin2\theta & \cos2\theta
    \end{pmatrix},
\end{equation}
where $\omega={\Delta m^2}/{4E}$ is the vacuum oscillation frequency with
$\Delta m^2$ being chosen as $ 2.0\times10^{-3}\operatorname{eV}^2 $,
$\sin^2\theta=10^{-4}$ is the mixing angle between two different flavors.

$H_{\rm mat} $ is induced by the interactions of neutrinos with the electrons, proton, or neutrons in the matter medium, which can be derived as
\end{multicols}
\begin{equation}
H_{\rm{mat}}(r)=\sqrt{2} G_{\mathrm{F}}  \operatorname{diag}\left(n_{e}(r)-\frac{n_{n}(r)}{2},-\frac{n_{n}(r)}{2},-n_{e}(r)+\frac{n_{n}(r)}{2}, \frac{n_{n}(r)}{2}\right)\,,
\end{equation}
\begin{multicols}{2}
where $n_{e}(r)$ and $n_{n}(r)$ are the number densities of electrons and neutrons at the radial position $r$, respectively. In our numerical calculation, we employ the
time dependent SN density profile in Ref.~\cite{Fogli:2007bk}, and the relation of $n_e=0.4 n_b$ is used~\cite{Duan:2006an}.

The third Hamiltonian component is from contributions of the neutrino self-interaction $H_{\rm self}$~\cite{deGouvea:2012hg},
\end{multicols}
\begin{equation}\label{H_self}
    H_{\rm self}=\mu(r)\int dEG\left(\rho(E)-\rho(E)^{\rm c*}\right)G+\frac{1}{2} G \operatorname{Tr}\left(\left(\rho(E)-\rho(E)^{\rm c*}\right) G\right),	
\end{equation}
\begin{multicols}{2}
where $\rho^{\rm c}$ is a transformation of $ \rho $ by replacing the $2\times2$ blocks of $ 11 \leftrightarrow 22 $ and $ 12 \leftrightarrow 21 $, and $ G $ is a diagonal matrix of dimensionless coupling constants which is defined as $ G=\operatorname{diag}\left(1,1,-1, -1\right) $.
$ \mu(r) $ is the strength of neutrino-neutrino interactions, which depends on the neutrino density profile and the geometric factor~\cite{Fogli:2007bk}
\begin{equation}
    \mu(r)=\sqrt{2} G_F [N_{\nu_{e}}(r)+N_{\bar{\nu}_e}(r)+N_{\nu_{x}}(r)+N_{\bar{\nu}_{x}}(r)]\,. 
\end{equation}

Finally, the Hamiltonian $H_{\rm mag}(r)$ accounting for the neutrino magnetic moment depends on the choice of Majorana neutrinos or Dirac neutrinos.
In this work, the Majorana nature of massive neutrinos has been assumed, and the Hamiltonian only has non-zero off-diagonal elements
\begin{equation}
H_{\rm mag}=\left(\begin{array}{cccc}0 & 0 & 0 & \mu_{\nu} B_T \\ 0 & 0 & -\mu_{\nu} B_T & 0 \\ 0 & -\mu_{\nu} B_T & 0 & 0 \\ \mu_{\nu} B_T & 0 & 0 & 0\end{array}\right),
\end{equation}
where $B_T$ is the magnetic field component transverse to the neutrino momentum $\mathbf{q}$, and $\mu_{\nu}$ is the neutrino magnetic moment. We assume a Standard-Model (SM) size magnetic moment as~\cite{Giunti:2014ixa,Giunti:2015gga}
\begin{equation}
 \mu_{\nu}({\rm SM})\simeq 3\times 10^{-19} \mu_{\rm B} \left(\frac{m_{\nu}}{1\;{\rm eV}}\right)\;,
\end{equation}
and a typical distribution of the SN magnetic field~\cite{Thompson:1993hn}
\begin{equation}
 B_T(r)\simeq 10^{12} \left(\frac{50\;{\rm km}}{r^2}\right)^2\;{\rm gauss}\;.
\end{equation}
Therefore the nominal value of $\mu_{\nu} B_T$ is taken as
\begin{eqnarray}
\left(\mu_{\nu} B_{T}\right)_{\operatorname{SM}}(r)\simeq
1.9 \times 10^{-10}\left(\frac{50 \mathrm{~km}}{r}\right)^{2}\left(\frac{\mathrm{eV}^{2}}{\mathrm{MeV}}\right)\;,
\end{eqnarray}
and other choices will be given in the unit of this nominal value.

Note that in order to fulfill the probability conservation, the density matrix satisfies the following relation
\begin{equation}
\int_{0}^{\infty} \operatorname{tr} \rho_{}(r) \mathrm{d} E=1\,,
\end{equation}
and the initial condition can be written as
\begin{equation}
    \rho^{\mathrm{ini}}(E)=\mathrm{diag}
    \begin{pmatrix}
        \frac{\phi_{\nu_e}(E)}{N\langle E_{\nu_e} \rangle}, \frac{\phi_{\nu_x}(E)}{N\langle E_{\nu_x} \rangle}, \frac{\phi_{\bar{\nu}_e}(E)}{N\langle E_{\bar{\nu}_e} \rangle}, \frac{\phi_{\bar{\nu}_x}(E)}{N\langle E_{\bar{\nu}_x} \rangle}
    \end{pmatrix}\,,
\end{equation}
where $N=\langle E_{\nu_e} \rangle^{-1}+\langle E_{\nu_x} \rangle^{-1}+\langle E_{\bar{\nu}_e} \rangle^{-1}+\langle E_{\bar{\nu}_x} \rangle^{-1}$ is the normalization factor.

\section{Spin Flavor Spectral Splits}

In this section, we present the results of numerically calculating of the neutrino flavor evolution equation in Eq.~\ref{evoltion equation}, and discuss the non-trivial phenomena of SN neutrino collective oscillations in the presence of neutrino magnetic moments and in the environment of strong magnetic field.
Our numerical calculation will be based on the approximations of two-flavor mixing and single-angle treatments.

\begin{figure}
    \centering
    \subfloat[]{
    \includegraphics[width=0.48\textwidth]{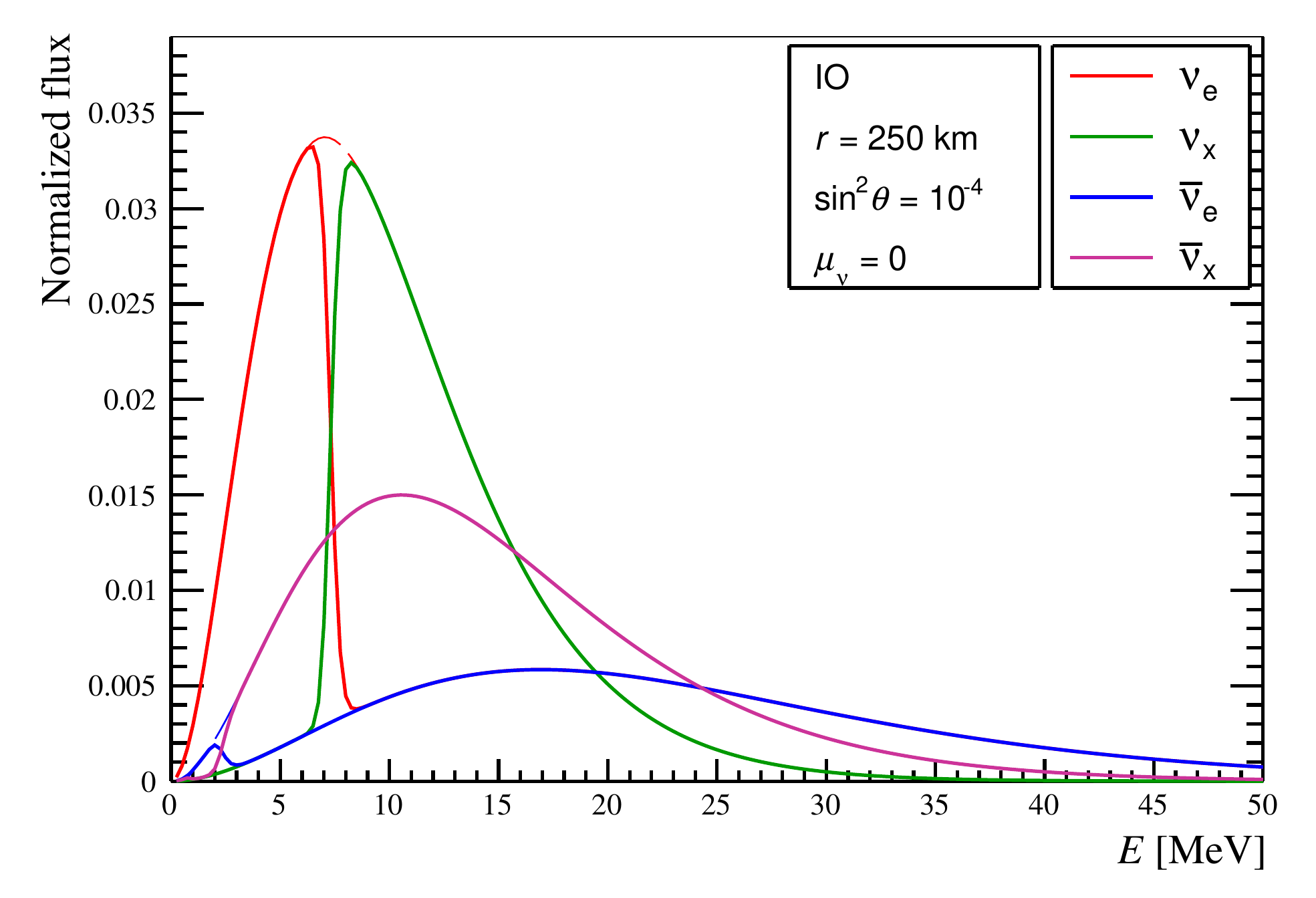} 
    }
    \subfloat[]{
    \includegraphics[width=0.48\textwidth]{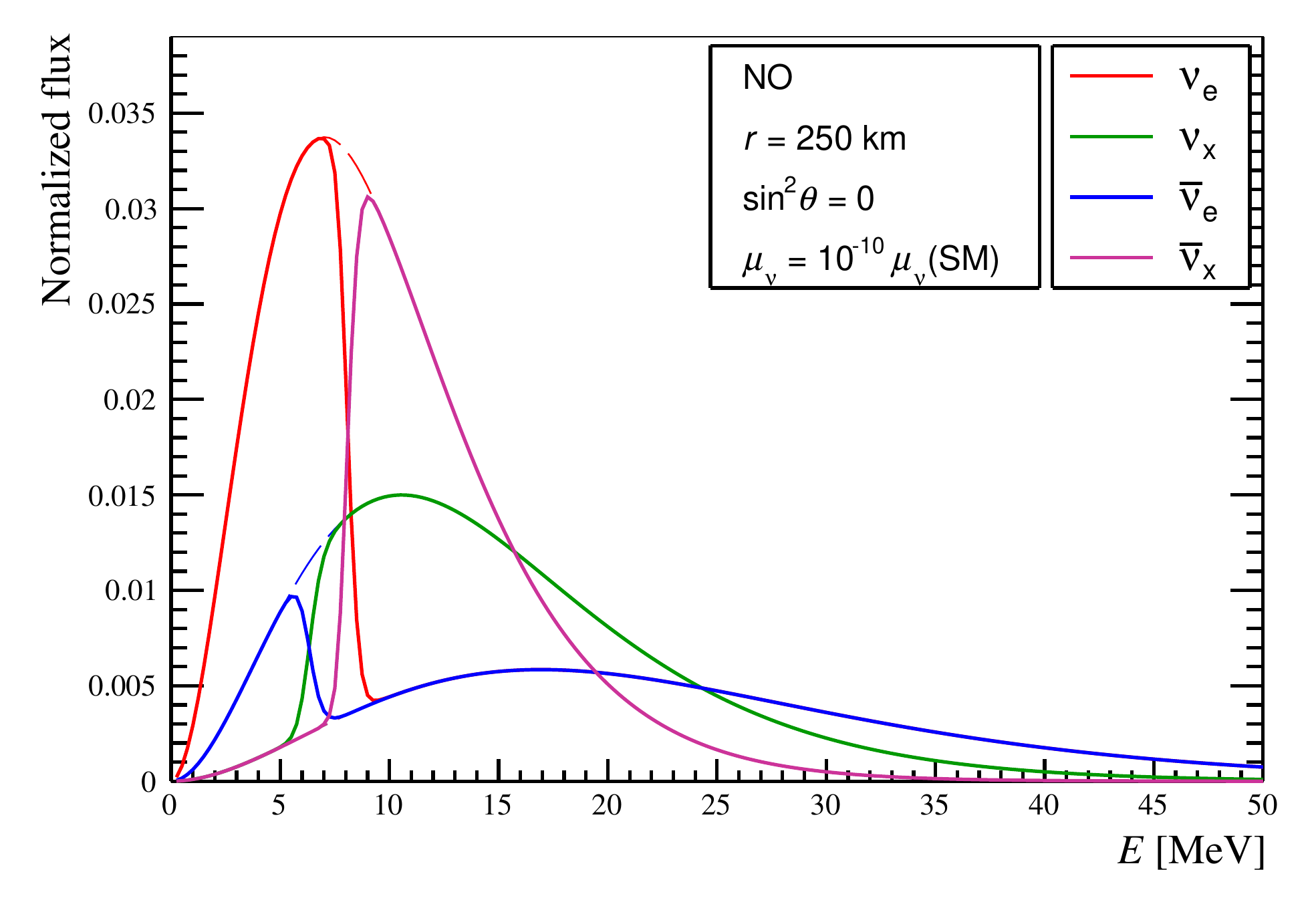} 
    }
    \figcaption{Spectral split feature of ($\nu_{e}$, $\nu_{x}$) and ($\overline{\nu}_{e}$, $\overline{\nu}_{x}$) for the inverted mass ordering by neglecting the neutrino magnetic moment ($\mu_{\nu}=0$) (left panel); and spectral split feature of ($\nu_{e}$, $\overline\nu_{x}$) and ($\overline{\nu}_{e}$, ${\nu}_{x}$) for the normal mass ordering neglecting the neutrino flavor mixing ($\theta=0$) (right panel).
    The dashed and solid lines are the neutrino energy spectra at $r=50$ km and $r=250$ km.}
    \label{Fig:compare}
\end{figure}

To understand the property of the collective oscillations induced by the neutrino magnetic moment better, we start with a comparison of two simplified cases with $\mu_{\nu}=0$ or $\theta=0$. The first case corresponds to the studies in Ref.~\cite{Fogli:2007bk}, in which the $4\times4$ neutrino flavor evolution equation will be reduced into two subsystems of ($\nu_{e}$, $\nu_{x}$) and ($\overline{\nu}_{e}$, $\overline{\nu}_{x}$) with two $2\times2$ evolution equations (see Eqs.~(34) and (35) in Ref.~\cite{Fogli:2007bk}). In the left panel of Fig.~\ref{Fig:compare}, we re-illustrate the spectral split feature between different flavor neutrinos for the inverted mass ordering, which illustrates almost complete spectral swaps at around 7 MeV for ($\nu_{e}$, $\nu_{x}$), and 2 MeV for ($\overline{\nu}_{e}$, $\overline{\nu}_{x}$). In contrast, for the second case of neglecting flavor mixing ($\theta=0$), the $4\times4$ neutrino flavor evolution equation will be reduced into two subsystems of ($\nu_{e}$, $\overline\nu_{x}$) and ($\overline{\nu}_{e}$, ${\nu}_{x}$) with two $2\times2$ evolution equations
\begin{equation}\label{pvtheta0}
\footnotesize
\begin{aligned}
\partial_{t} \mathbf{P}^+ &=[+\omega \mathbf{B_{\rm eff}}+\mu(-P^+_{x}+P^-_{x},-P^+_{y}+P^-_{y},P^+_{z}+P^-_{z})] \times \mathbf{P}^+ \\
\partial_{t} {\mathbf{P}}^- &=[-\omega \mathbf{B_{\rm eff}}+\mu(-P^-_{x}+P^+_{x},-P^-_{y}+P^+_{y},P^-_{z}+P^+_{z})] \times {\mathbf{P}}^-
\end{aligned}
\footnotesize
\end{equation}
where
$\mathbf{B_{\rm eff}} = (\sin2\theta_{\rm eff},\; 0,\; -\cos2\theta_{\rm eff})$, $\tan{2\theta_{\rm eff}}={\mu_{\nu} B_{T}}/{\omega}$,
and
the new polarisation vectors $\mathbf{P}^+$ and $\mathbf{P}^-$ are defined as
\begin{equation}
    \rho^+=\left(\begin{array}{cc} \rho_{ee} & \rho_{e\bar{x}} \\ \rho_{\bar{x}e} & \rho_{\bar{x}\bar{x}}
    \end{array}\right)=\frac{1}{2}(1+\mathbf{P}^+\cdot\mathbf{\sigma})\,,
\end{equation}
and
\begin{equation}
    \rho^-=\left(\begin{array}{cc} \rho_{xx} & \rho_{x\bar{e}} \\ \rho_{\bar{e}x} & \rho_{\bar{e}\bar{e}}
    \end{array}\right)=\frac{1}{2}(1+\mathbf{P}^-\cdot\mathbf{\sigma})\,.
\end{equation}
The neutrino energy spectra of different flavors at $r=50$ km and $r=250$ km from the numerical solution of the evolution equations of Eq.~\ref{pvtheta0} are illustrated in the right panel of Fig.~\ref{Fig:compare} for the normal mass ordering,
where the spin flavor spectral split appears between $\nu_{e}$ and $\overline\nu_{x}$ and between $\overline{\nu}_{e}$ and ${\nu}_{x}$, respectively.
The location of spectral swaps varies when one changes the value of the neutrino magnetic moment.
If $\mu_{\nu}$ is taken as $10^{-10}\mu_{\nu}({\rm SM})$ the swap location will be at around {8} MeV for ($\nu_{e}$, $\overline\nu_{x}$), and 6 MeV for ($\overline{\nu}_{e}$, $\nu_{x}$), qualitatively similar to those of usual flavor spectral splits in the case of $\mu_{\nu}=0$ and $\theta\neq0$, which can be understood by the essential correspondence between Eq.~\ref{pvtheta0} and Eqs.~(34-35) in Ref.~\cite{Fogli:2007bk}.

\end{multicols}
\begin{multicols}{2}
These two kinds of neutrino energy spectral splits fall into the category of the so-called single split type~\cite{Fogli:2007bk},
in which the detailed analysis has been discussed. For the left panel in Fig.~\ref{Fig:compare},
the critical energies can be estimated by requiring the lepton number conversion:
\begin{equation}
\int^{\infty}_{E_1}(n_{{\nu}_e}-n_{\nu_x})=\int^{\infty}_{E_2}(n_{\bar{\nu}_e}-n_{\bar{\nu}_x} )\,, \label{eq:single}
\end{equation}
where $E_1$ and $E_2$ depend on both the mass-mixing parameters and the neutrino energy spectra, and according to the current inputs, one can obtain that
$E_1\simeq7$ MeV and $E_2\simeq2$ MeV.
As for the right panel with the non-zero neutrino magnetic moment, an effective lepton number conservation of each sub-system can be defined as
\begin{equation}
\int^{\infty}_{E_1}(n_{{\nu}_e}-n_{\bar{\nu}_x})=\int^{\infty}_{E_2}(n_{\bar{\nu}_e}-n_{{\nu}_x} ).\label{eq:single2}
\end{equation}
where $E_1$ and $E_2$ can be similarly calculated by using the neutrino energy spectra and the effective mixing $\theta_{\rm eff}$.
In summary, we conclude that the neutrino magnetic moment is playing the role of effective mixing in the subsystems of ($\nu_{e}$, $\overline\nu_{x}$) and ($\overline{\nu}_{e}$, ${\nu}_{x}$), which provides an initial perturbation for the transition between different spin states (i.e., neutrino or antineutrino), and when the neutrino self-interaction Hamiltonian is included, it will trigger non-trivial spin flavor neutrino collective oscillations as shown in the right panel of Fig.~\ref{Fig:compare}. In the following we shall present the numerical calculations in the general case of
$\mu_{\nu}\neq0$ and $\theta\neq0$.

\end{multicols}
\begin{figure}
    \centering
    \subfloat[]{
    \includegraphics[width=0.48\textwidth]{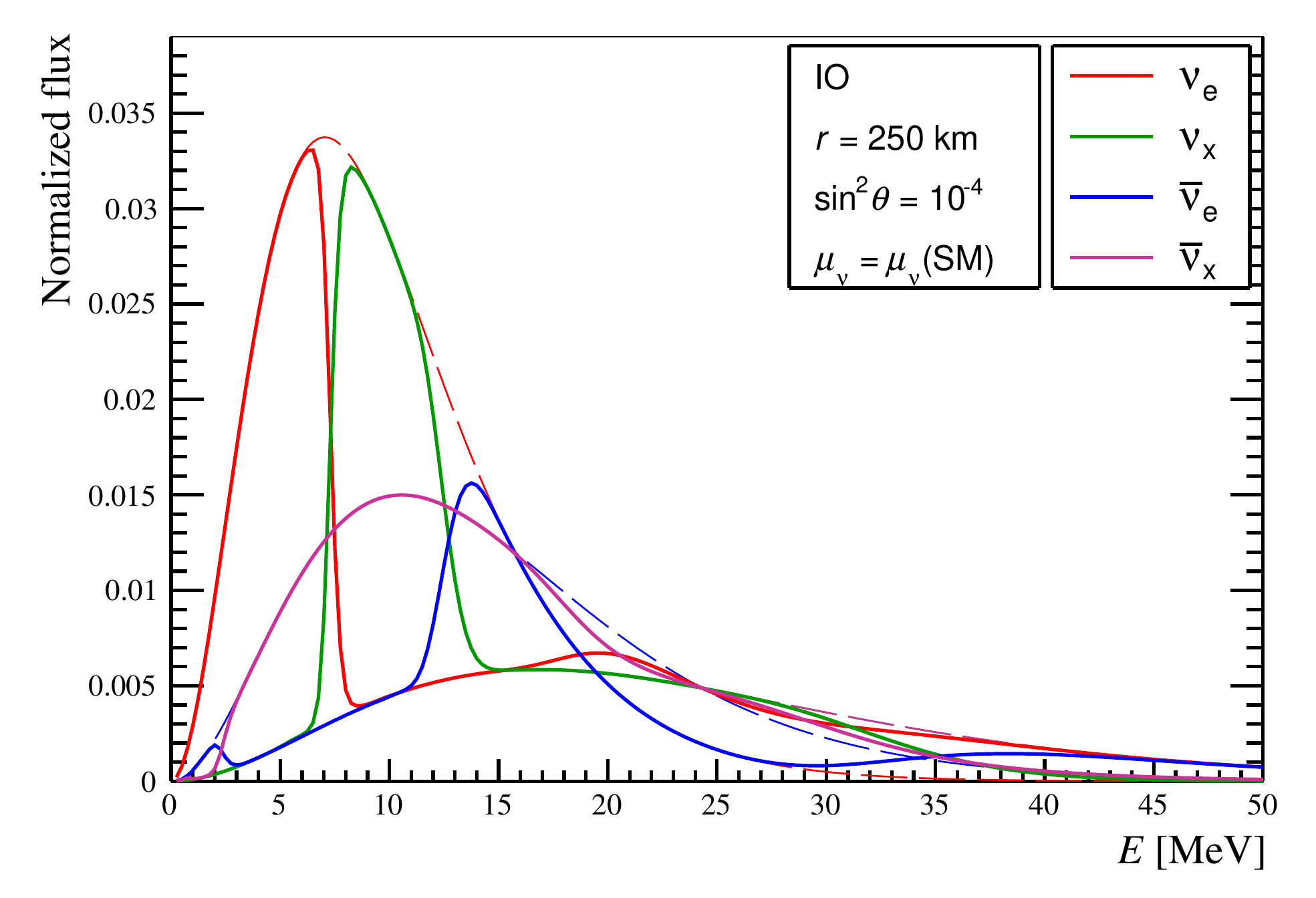} 
    }
    \subfloat[]{
    \includegraphics[width=0.48\textwidth]{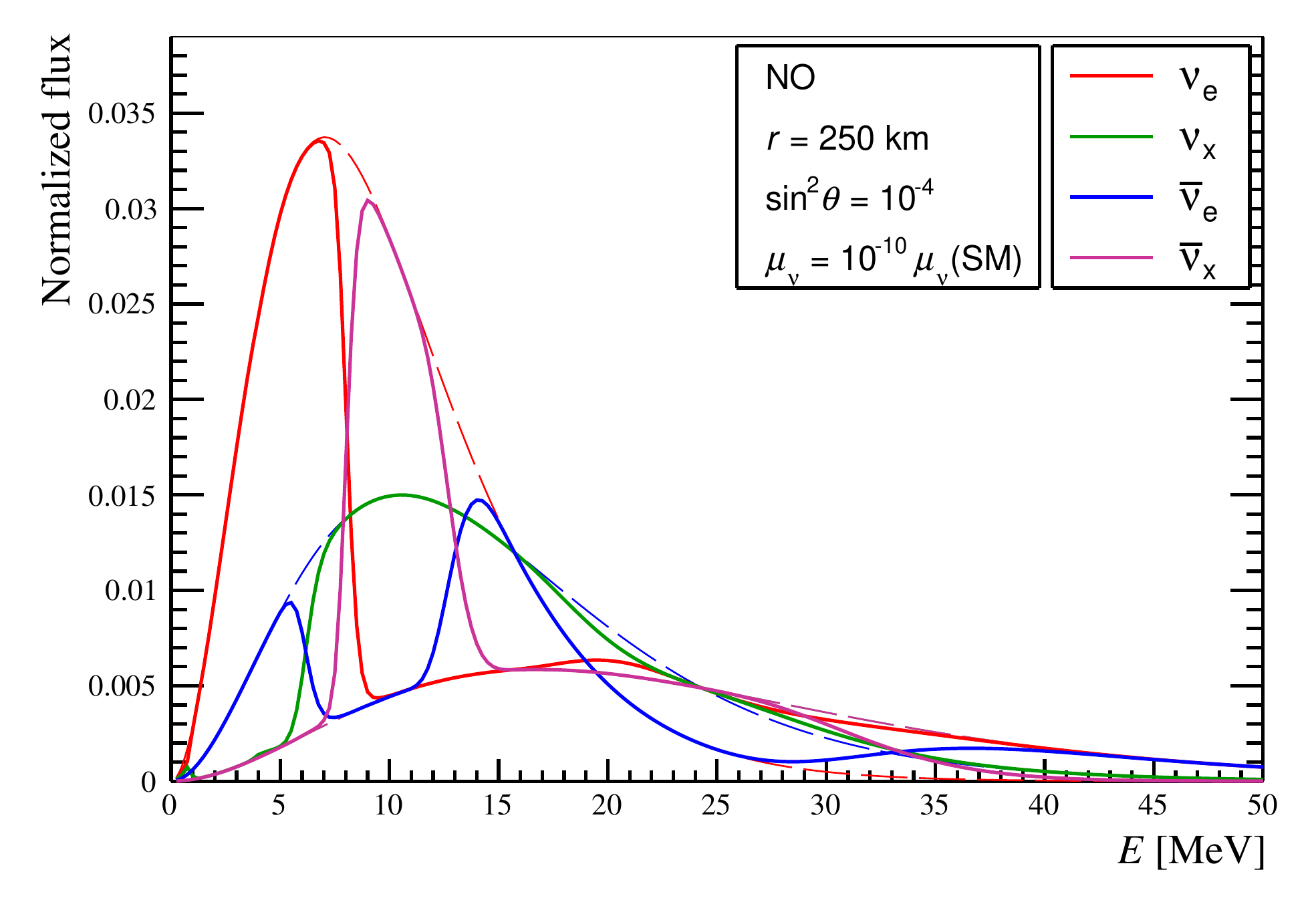} 
    }
    \figcaption{Spectral split feature of the ($\nu_{e}$, $\nu_{x}$, $\overline{\nu}_{e}$, $\overline{\nu}_{x}$) framework the inverted mass ordering with $\mu_{\nu}=\mu_{\nu}(\rm SM)$ (left panel) and normal mass ordering with $\mu_{\nu}=10^{-10}\mu_{\nu}(\rm SM)$ (right panel).
    The dashed and solid lines are the neutrino energy spectra at $r=50$ km and $r=250$ km.}
    \label{Fig:4flavorsplit}
\end{figure}
\begin{multicols}{2}
In Fig.~\ref{Fig:4flavorsplit}, we present the calculation results of the spectral split feature in the ($\nu_{e}$, $\nu_{x}$, $\overline{\nu}_{e}$, $\overline{\nu}_{x}$) framework for the inverted mass ordering with $\mu_{\nu}=\mu_{\nu}(\rm SM)$ (left panel) and normal mass ordering with $\mu_{\nu}=10^{-10}\mu_{\nu}(\rm SM)$ (right panel).
The vacuum mixing parameter is taken as $\sin^2\theta=10^{-4}$. The dashed and solid lines are the neutrino energy spectra at $r=50$ km and $r=250$ km.

From the figure, one can observe that the spectral split feature is rather complicated in the presence of both the flavor mixing and neutrino magnetic moment. These splits can be classified into two parts. For the spectra below 10 MeV, the split behavior is rather similar to those of simplified scenarios in Fig.~\ref{Fig:compare}.
For the inverted ordering, the single split between different flavors in the left panel of Fig.~\ref{Fig:compare} is generally reserved in the left panel of Fig.~\ref{Fig:4flavorsplit} when the neutrino magnetic moment is included. Meanwhile the single split between the different spin-flavor states in the right panel of Fig.~\ref{Fig:compare} for the normal ordering case is also kept in the right panel of Fig.~\ref{Fig:4flavorsplit} when we consider the flavor mixing, indicating the fact that one effect is dominating over the other for the neutrino energies below 10 MeV. Next let us take a closer look at the spectral splits above 10 MeV.
Interestingly, spin-flavor spectral split appears for the inverted ordering, but flavor spectral split happens for the normal ordering. However, the split behavior is completely different, which belongs to the type of the multiple spectral split discussed in Ref.~\cite{Dasgupta:2009mg}. Taking the ($\overline{\nu}_{e}$, $\nu_{x}$) pair in the left panel of Fig.~\ref{Fig:4flavorsplit} as an example, they first swap their spectra at the location of $E_{1}\simeq12$ MeV, and finally switch back at around $E_{3}\simeq35$ MeV. In between the spectra naturally cross each other at $E_{2}\simeq19$ MeV. The location of the multiple spectral splits can also be inferred from the spirit of an effective lepton number conservation:
\begin{equation}\label{eq:multi}
    \begin{aligned}
    \int^{E_3}_{E_1}(n_{\overline{\nu}_e}-n_{\nu_x})=\int^{E_2}_{E_1}(n_{\overline{\nu}_e}-n_{\nu_x})-\int^{E_{3}}_{E_2} (n_{\nu_x}-n_{\overline{\nu}_e})=0\,,
    \end{aligned}
\end{equation}
which depends on both the neutrino spectra and the physical parameters of flavor mixing and the neutrino magnetic moment.
Similar observation and analyses can also be applied to the spin flavor spectral split of the (${\nu}_{e}$, $\overline{\nu}_{x}$) pair in the left panel of Fig.~\ref{Fig:4flavorsplit}, and the flavor spectral split of the (${\nu}_{e}$, ${\nu}_{x}$) and ($\overline{\nu}_{e}$, $\overline{\nu}_{x}$) pairs in the right panel of Fig.~\ref{Fig:4flavorsplit}, where different relations of the lepton number conservation can be used to understand the properties of these splits.

Note that the multiple spectral split occurs within a sub-system of two different neutrino flavors, but the single splits in Fig.~\ref{Fig:compare} involve all the four neutrinos. The single split can be regarded as the special cases of two multiple splits. Because the energy spectra of different neutrinos are rather different, there are possible scenarios that Eq.~\ref{eq:multi} cannot be fulfilled even when $E_{3}$ goes to infinity. Therefore other neutrino flavors needs to participate in the conversion to keep the lepton number conservation as shown in Eq.~\ref{eq:single2}, which is in the type of the single spectral split.
These qualitative analyses can be achieved in the presence of the nonlinear neutrino self-interactions and the property obtained belongs to the general behavior of neutrino collective oscillations.

\end{multicols}
\begin{figure}
    \centering
    \subfloat[]{
    \includegraphics[width=0.45\textwidth]{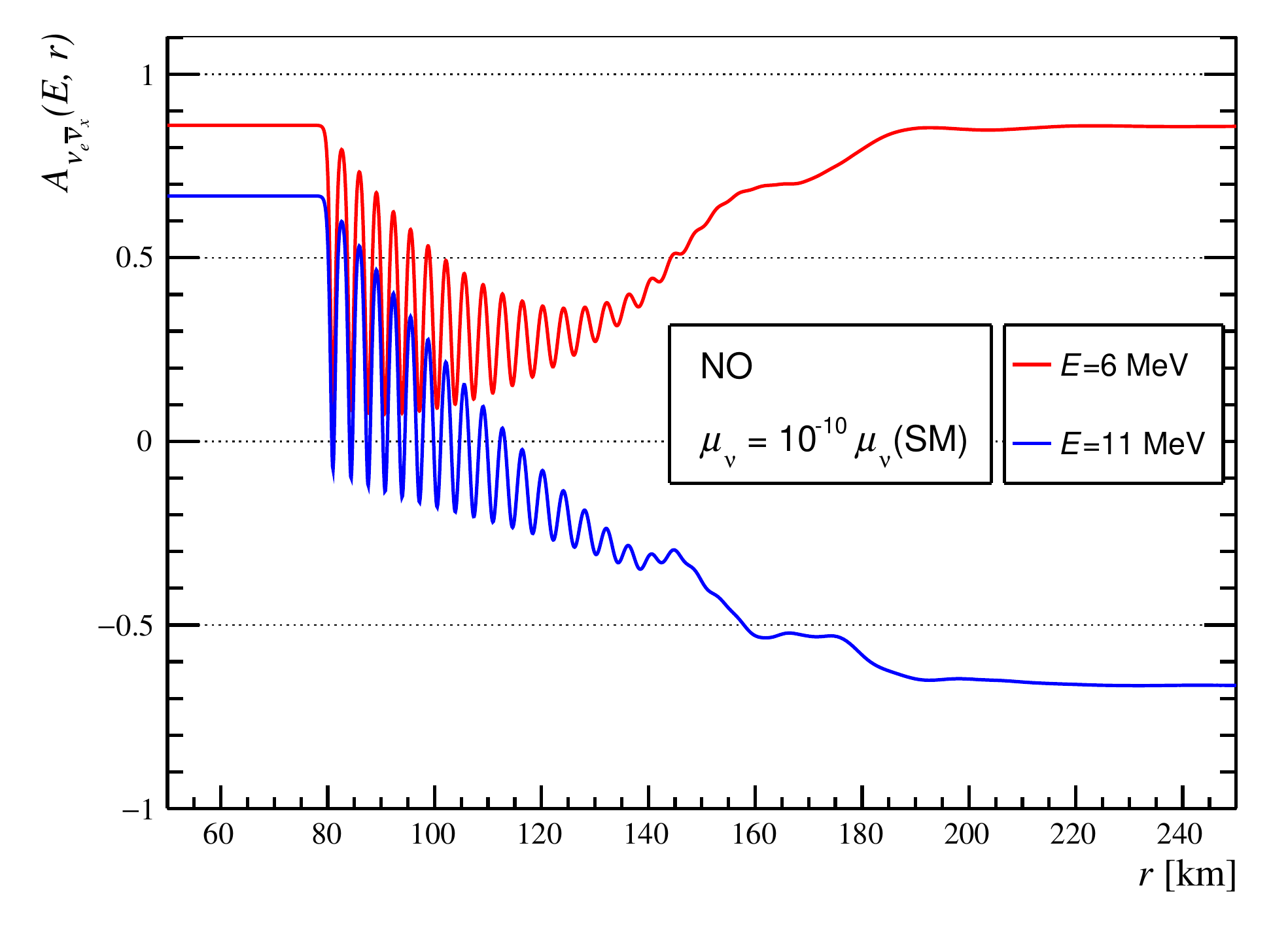} 
    }
    \subfloat[]{
    \includegraphics[width=0.45\textwidth]{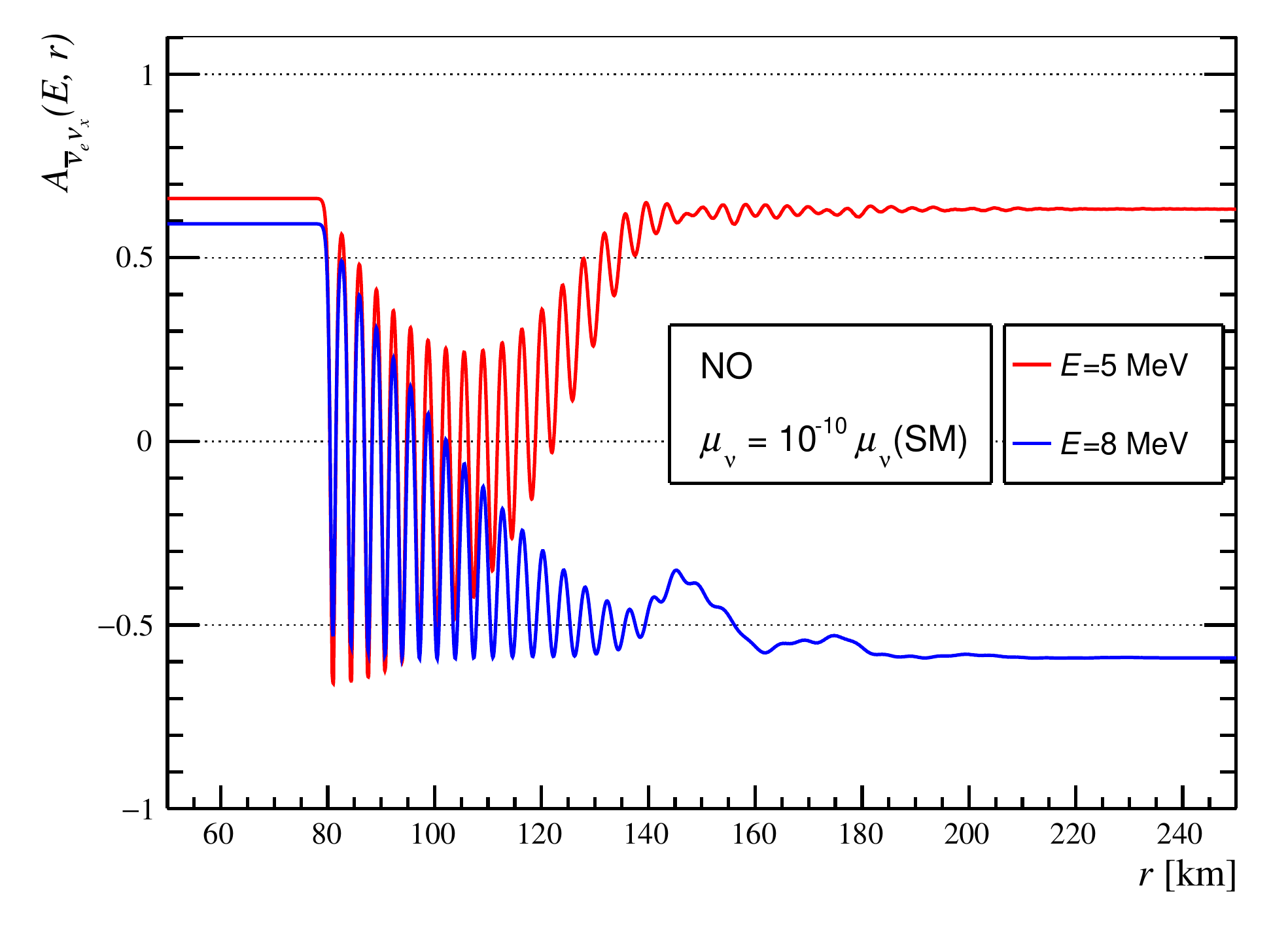}
    }\\
    \subfloat[]{
    \includegraphics[width=0.45\textwidth]{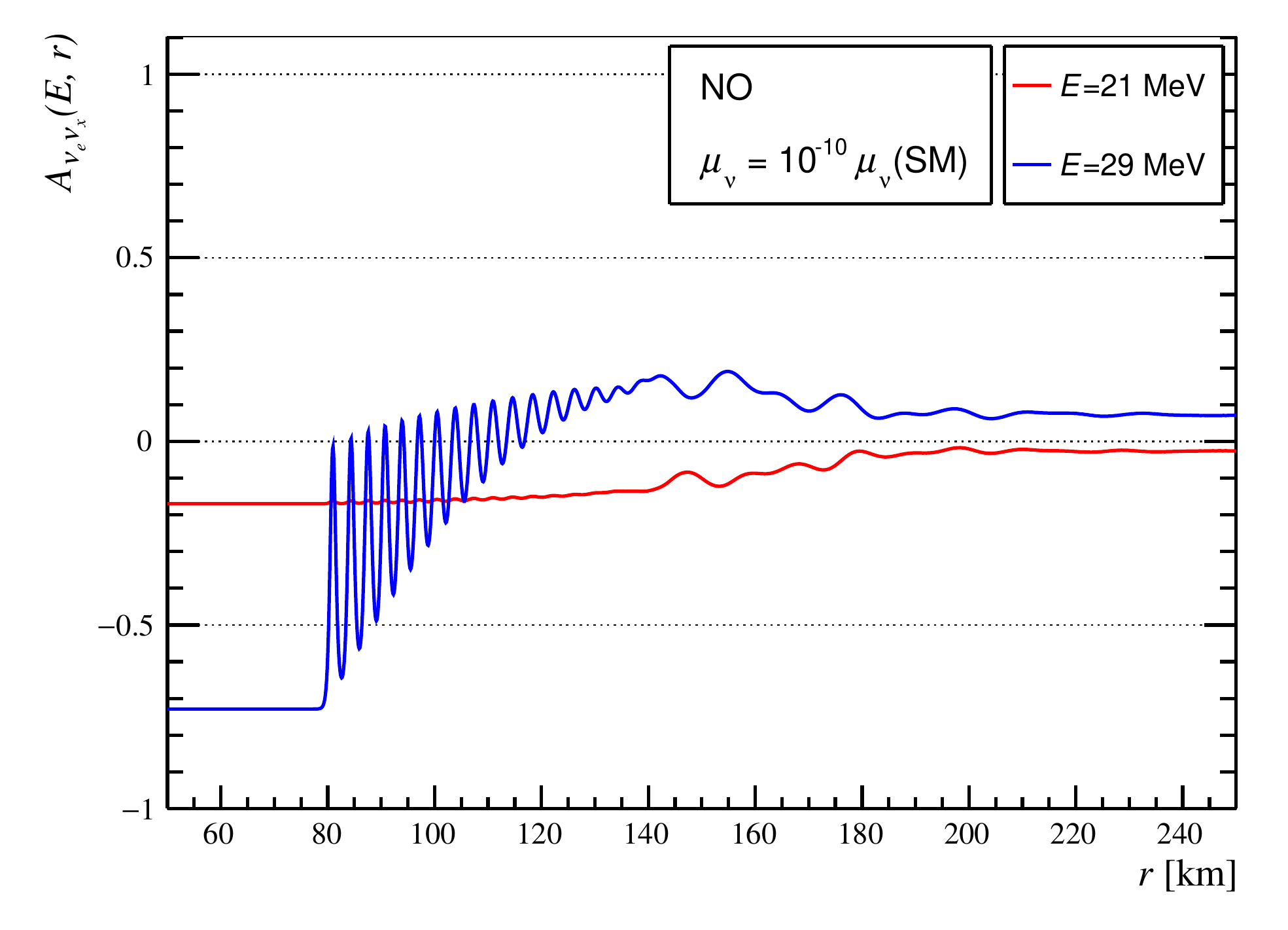}
    }
    \subfloat[]{
    \includegraphics[width=0.45\textwidth]{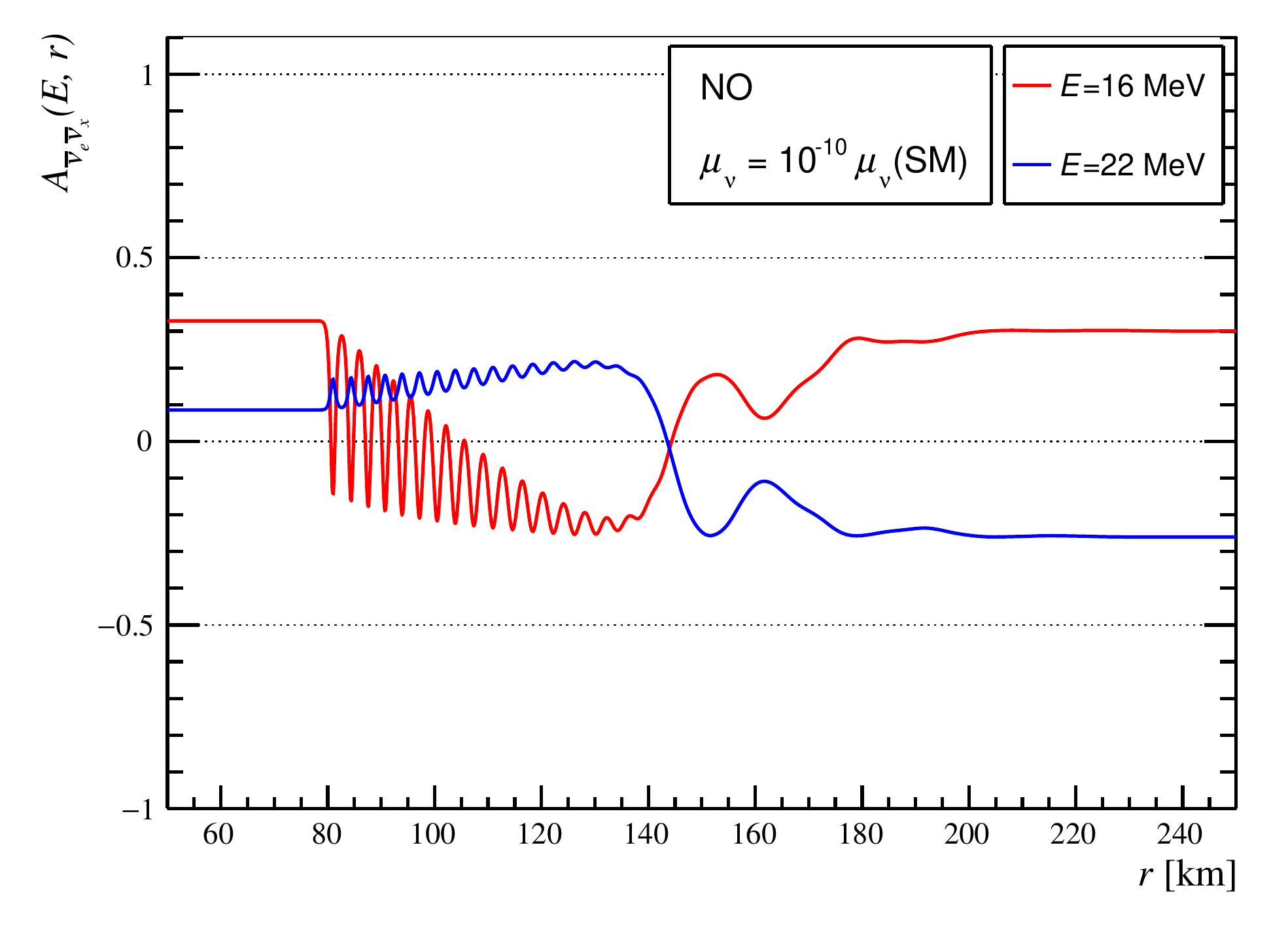}
    }
    \figcaption{Flux asymmetries $A_{{\nu_{\alpha}}{\nu_{\beta}}}(E,r)$ as functions of the neutrino positions for several typical neutrino energies, where the upper left, upper right, lower left and lower right panels are for the pairs of (${\nu}_{e}$, $\overline{\nu}_{x}$), ($\overline{\nu}_{e}$, ${\nu}_{x}$), (${\nu}_{e}$, ${\nu}_{x}$), and ($\overline{\nu}_{e}$, $\overline{\nu}_{x}$) respectively.}
    \label{Fig:positon}
\end{figure}
\begin{multicols}{2}

Next let us study how the spectral split forms from the neutrino sphere to the position of $r=250$ km.
We first define the flux asymmetries $A_{{\nu_{\alpha}}{\nu_{\beta}}}(E,r)$ as functions of the neutrino energy and position for different neutrino flavors
\begin{equation}
    A_{{\nu_{\alpha}}{\nu_{\beta}}}(E,r)=\frac{n_{\nu_{\alpha}}(E,r)-n_{\nu_{\beta}}(E,r)}{n_{\nu_{\alpha}}(E,r)+n_{\nu_{\beta}}(E,r)}\;.
\end{equation}
Taking the normal mass ordering with $\mu_{\nu}=10^{-10}\mu_{\nu}(\rm SM)$ as an example, the flux asymmetries in several typical neutrino energies
are shown in Fig.~\ref{Fig:positon}, in which the upper left, upper right, lower left and lower right panels are for the pairs (${\nu}_{e}$, $\overline{\nu}_{x}$), ($\overline{\nu}_{e}$, ${\nu}_{x}$), (${\nu}_{e}$, ${\nu}_{x}$), and ($\overline{\nu}_{e}$, $\overline{\nu}_{x}$) respectively.

From the figure, we find that the flavor evolution from $r=50$ km to $r=250$ km can be divided into four different phases. The first phase is the synchronized oscillation~\cite{Pastor:2001iu}, which happens between 50 km and 75 km, where all the neutrinos and antineutrinos of different energies oscillate with the same frequency and tiny amplitude, thus the neutrino number density will keep unchanged.
The second phase, starting from 75 km to 140 km in the upper left and upper right panels, is the bipolar oscillation~\cite{Hannestad:2006nj} induced by the neutrino magnetic moment. This bipolar oscillation is responsible for the single spectral splits below 10 MeV between the spin-flavor states of (${\nu}_{e}$, $\overline{\nu}_{x}$), ($\overline{\nu}_{e}$, ${\nu}_{x}$) pairs.
The third phase happens between 140 km and 180 km, as shown in the lower left and lower right panels of (${\nu}_{e}$, ${\nu}_{x}$), and ($\overline{\nu}_{e}$, $\overline{\nu}_{x}$) pairs, where a second favor-mixing induced bipolar oscillation takes place with rather slower frequencies, and is responsible for the multiple spectral splits above 10 MeV.
Note that the flavor-mixing induced bipolar oscillation is absent for the normal mass ordering in Ref.~\cite{Fogli:2007bk}, but arises in this work.
In the presence of the neutrino magnetic moment, the new spin-flavor bipolar oscillations of ($\nu_e$,\;$\overline{\nu}_x$) and ($\nu_x$,\;$\overline{\nu}_e$)
can significantly alter the energy spectra of different flavor neutrinos. Therefore the $\overline{\nu}_x$ flux would dominate over $\nu_e$, and the condition for the spectral swap can be satisfied for the neutrino energies above 10 MeV.
Finally the fourth phase occurs after 180 km, where the ordinary vacuum oscillation is expected. However due to the tiny flavor mixing, the flavor conversion can be neglected. In summary, after all the four phases, we can obtain the patterns of the flavor and spin-flavor spectral splits as shown in the right panel of Fig.~\ref{Fig:4flavorsplit}.

\end{multicols}
\begin{figure}
    \centering
    \subfloat[]{
    \includegraphics[width=0.45\textwidth]{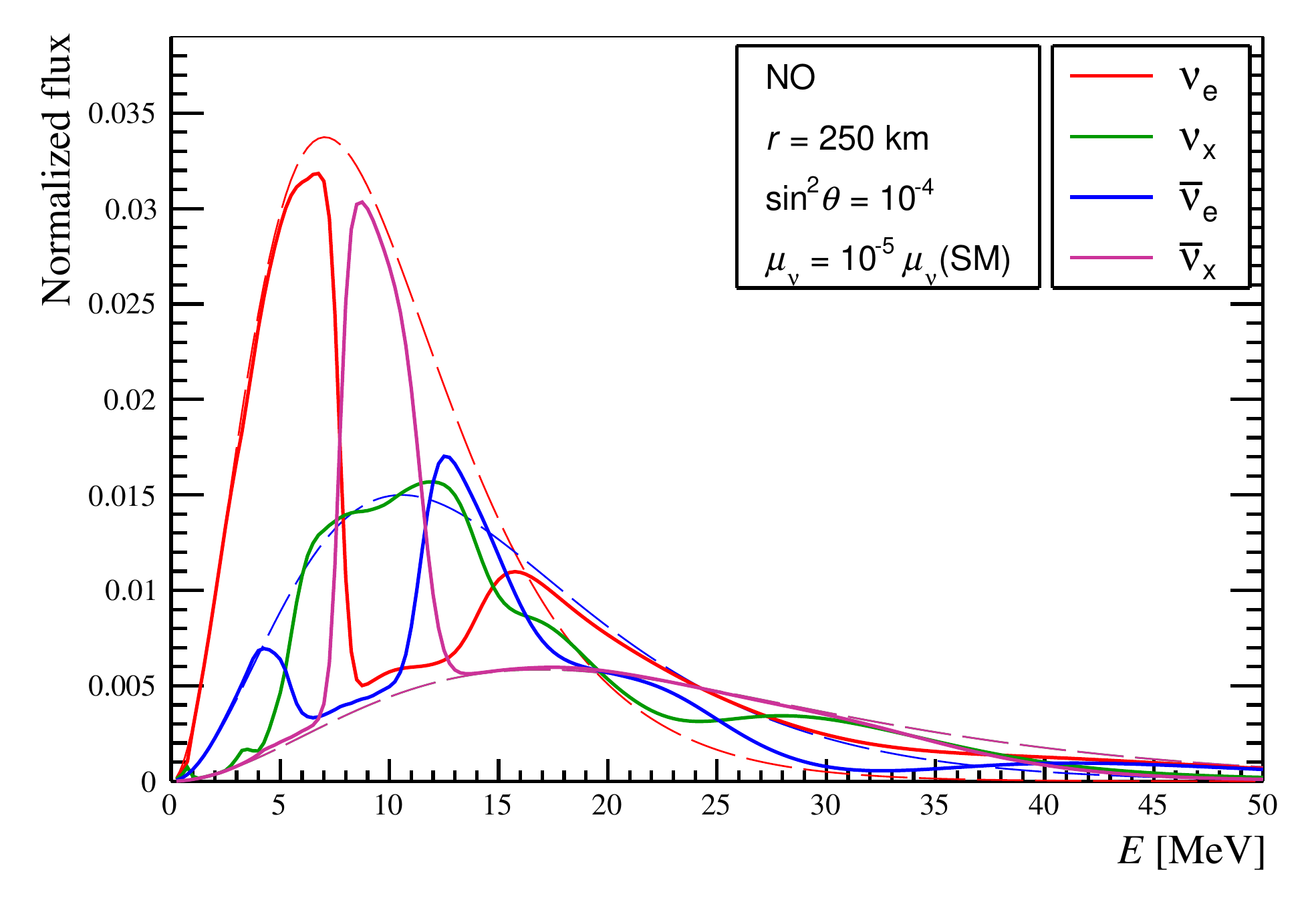} 
    }
    \subfloat[]{
    \includegraphics[width=0.45\textwidth]{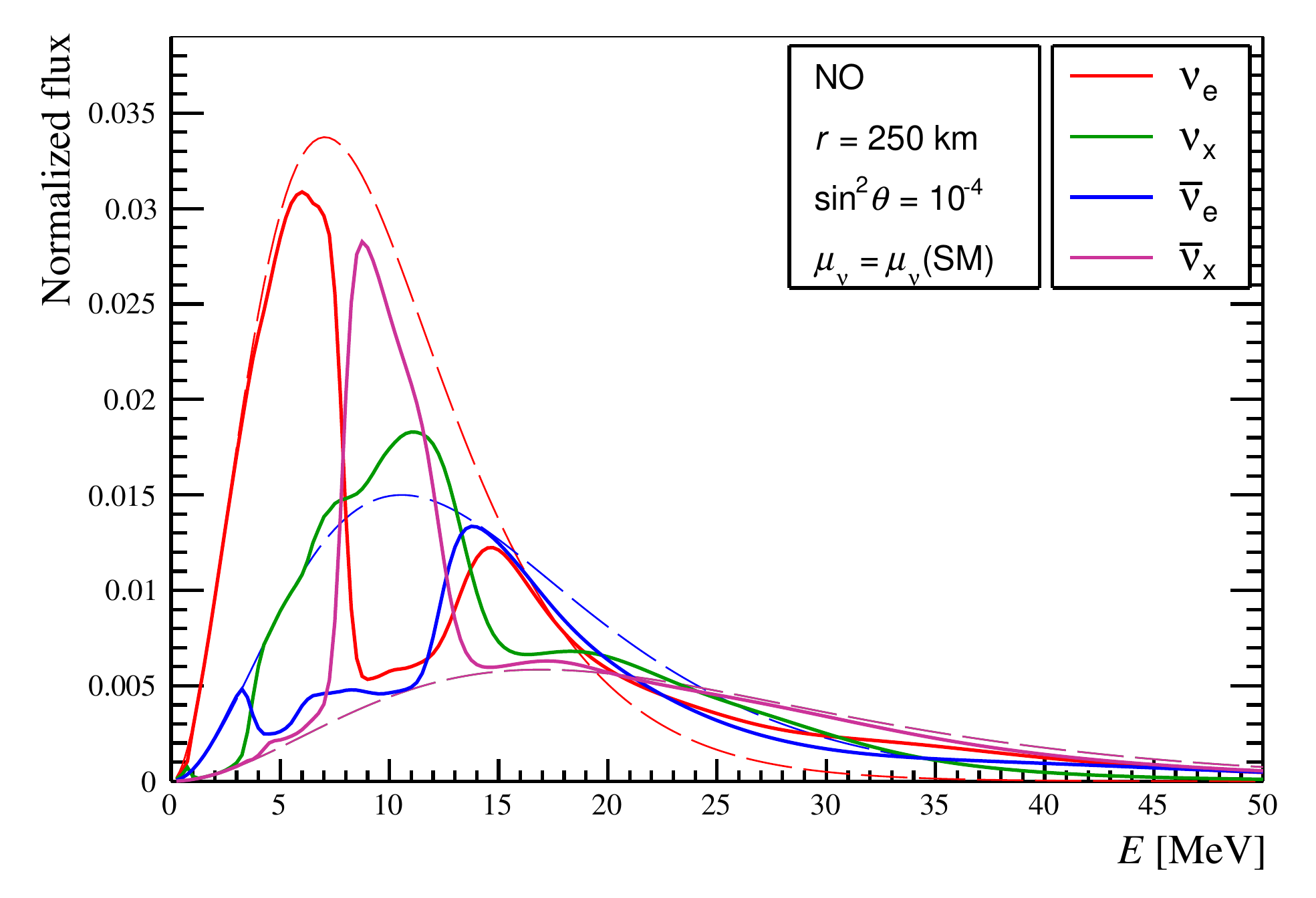} 
    }\\
    \subfloat[]{
    \includegraphics[width=0.45\textwidth]{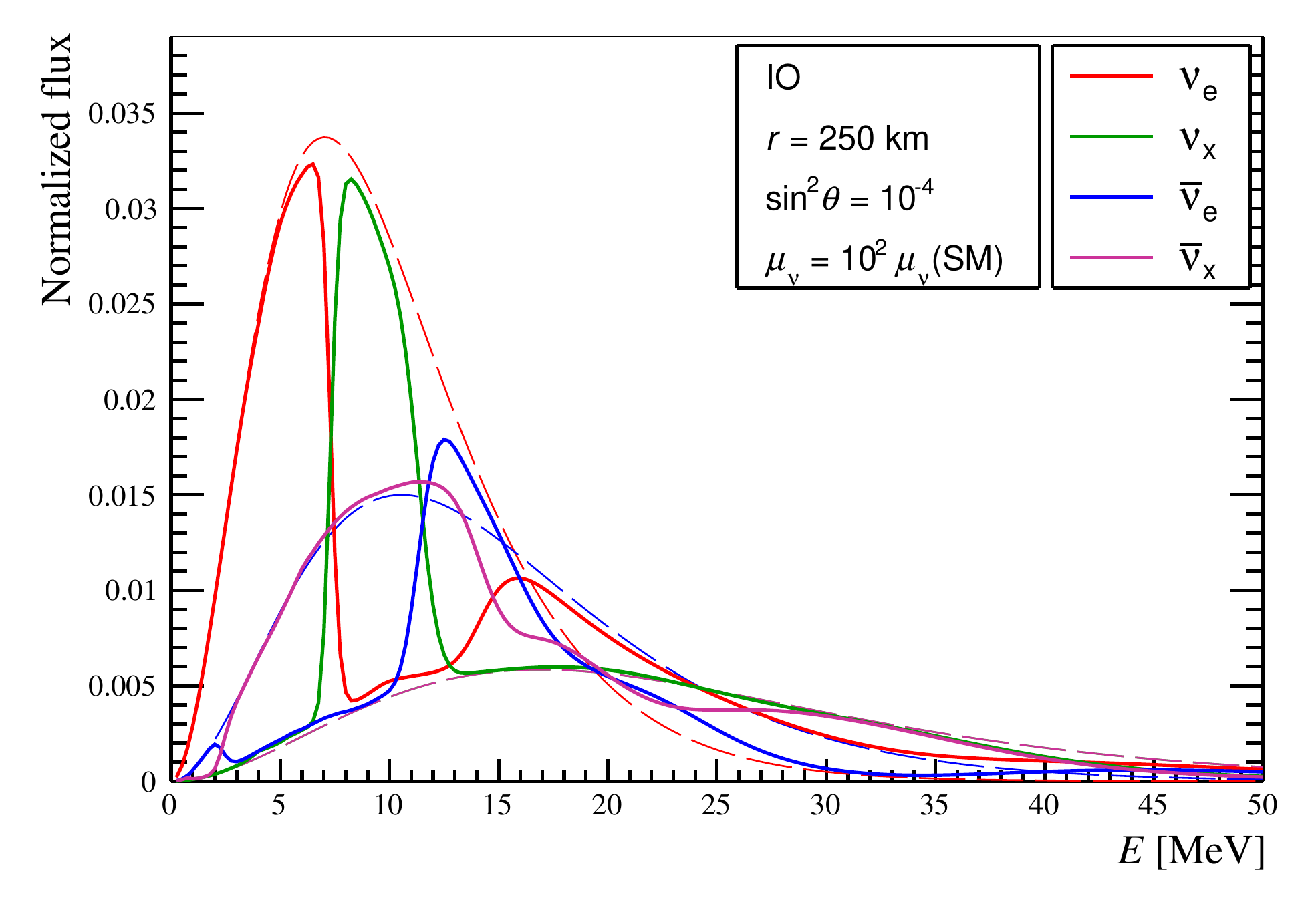}
    }
    \subfloat[]{
    \includegraphics[width=0.45\textwidth]{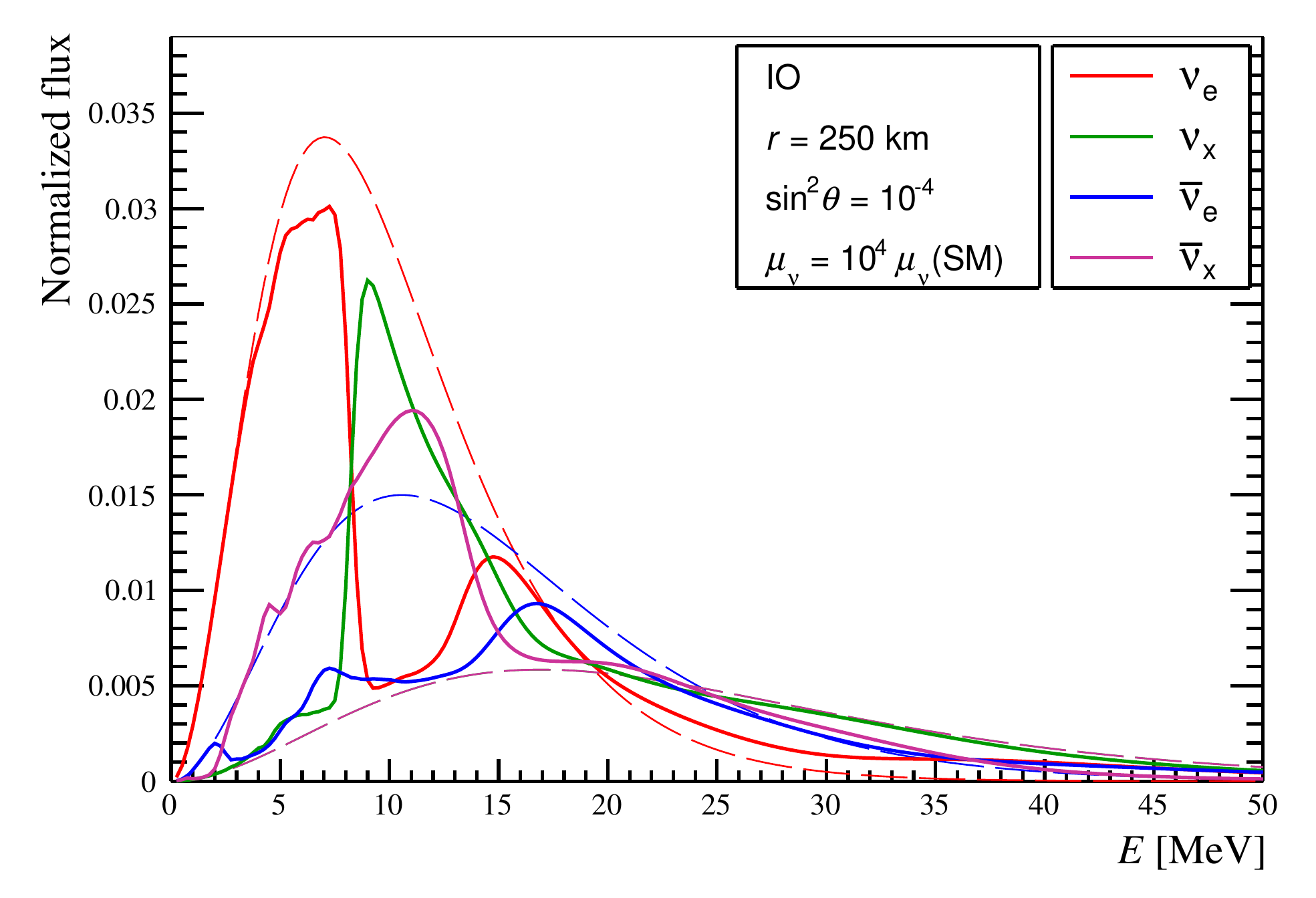}
    }
    \figcaption{Spectral split feature of the ($\nu_{e}$, $\nu_{x}$, $\overline{\nu}_{e}$, $\overline{\nu}_{x}$) framework for the normal mass ordering (upper panels) with $\mu_{\nu}=10^{-5}\;\mu_{\nu}(\rm SM)$ (left) and $\mu_{\nu}=\mu_{\nu}(\rm SM)$ (right), and for the inverted mass ordering (lower panels) with $\mu_{\nu}=10^{2}\;\mu_{\nu}(\rm SM)$ (left) and $\mu_{\nu}=10^{4}\;\mu_{\nu}(\rm SM)$ (right).
    The dashed and solid lines are the neutrino energy spectra at $r=50$ km and $r=250$ km.}
    \label{Fig:size}
\end{figure}
\begin{multicols}{2}
Before finishing this section, we would like to discuss how the size of the neutrino magnetic moment changes the patterns of spectral splits.
Taking the normal (inverted) mass ordering and enlarging the magnitude of the magnetic moment to $\mu_{\nu}=10^{-5}\mu_{\nu}(\rm SM)$ ($\mu_{\nu}=10^{2}\mu_{\nu}(\rm SM)$) and $\mu_{\nu}=\mu_{\nu}(\rm SM)$ ($\mu_{\nu}=10^{4}\mu_{\nu}(\rm SM)$),
we illustrate in Fig.~\ref{Fig:size} the spectral split feature of the neutrino flavor and spin-flavor collective oscillations in the ($\nu_{e}$, $\nu_{x}$, $\overline{\nu}_{e}$, $\overline{\nu}_{x}$) framework.
When the magnetic moment is small, as shown in both panels of Fig.~\ref{Fig:4flavorsplit},
each kind of spectral splits takes place within a rather separated energy and spatial region. However, as the magnetic moment gets larger, the pattern of spectral
splits would become complicated and look rather messy in Fig.~\ref{Fig:size}, where two successive bipolar oscillations induced by the
magnetic moment and flavor mixing would be mixed together and there will be no single complete spectral split. However one can still recognize the single spectral splits below 10 MeV. Comparing the situations in the normal and inverted mass orderings, one can learn that the neutrino magnetic moment is much more sensitive in the former case, and even that much smaller than the SM prediction, could have significant effects on the neutrino spectra, and may be observable in the future SN neutrino observation.

\section{Conclusion}

In this work, we have performed a numerical calculation of the neutrino flavor evolution inside the SN medium in the presence of the neutrino magnetic moment and in the environment of the strong magnetic field. By using the two-flavor and single-angle approximation, we first demonstrate that the neutrino magnetic moment can serve as an effective mixing between different neutrino spin-flavor states. We observe both flavor and spin-flavor neutrino collective oscillations inside the dense SN medium.
Fruitful neutrino spectral splits have been identified.
In the low energy range below around 10 MeV, single spectral split between different flavor states occurs for the case of inverted mass ordering, while the single spectral split between different spin-flavor states takes place for the case of normal mass ordering. For the energy range above 10 MeV, we observe multiple spectral splits within sub-systems of two different neutrinos, either pure flavor conversions (normal ordering) or spin-flavor conversions (inverted ordering).
We also find that the pattern of the flavor and spin-flavor spectral splits strongly depends on the magnitude of the neutrino magnetic moment, and even if the magnetic moment is smaller than the SM prediction, there will be significant effects on the SN neutrino spectra.

The results in this study are calculated within the two-flavor and single-angle approximation. Including the effects of the three-flavor mixing and multi-angle calculations would make the calculation more complicated and rather time-consuming. According to the experience of the pure flavor neutrino collective oscillations, the general features will preserve, but some new features may emerge~\cite{Dasgupta:2007ws,Duan:2007sh,Cherry:2010yc}. Therefore, we anticipate it will be similar cases for the spin-flavor collective oscillations, which will be reported in a future separated work.

 \medskip

\begin{acknowledgments}

The authors are grateful to Alexander Studenikin, Zhenyu Zhang and Shun Zhou for the helpful discussions.
This work is supported by National Natural Science Foundation of China under Grant Nos.~11835013, 12075255, and 11390381. Y.F. Li is also grateful for the support by the CAS Center for Excellence in Particle Physics (CCEPP).

\medskip


\end{acknowledgments}

\end{multicols}

\end{document}